\documentstyle[prl,aps]{revtex}
\begin{document}
\twocolumn[\hsize\textwidth\columnwidth\hsize\csname %
@twocolumnfalse\endcsname

\draft
\preprint{CTP-2658}
\title{The Square-Lattice Heisenberg Antiferromagnet at Very Large
Correlation Lengths}
\author{B.B. Beard$^a$, R.J. Birgeneau$^b$, M. Greven$^b$,
and U.-J. Wiese$^b$  \\
$^a$ Departments of Physics and Mechanical Engineering, Christian Brothers
University, Memphis, TN 38104 \\
$^b$ Department of Physics, Massachusetts Institute of Technology,
Cambridge, MA 02139 \\
}
\date{\today}
\maketitle
\begin{abstract}

The correlation length of the square-lattice spin-1/2 Heisenberg
antiferromagnet is studied in the low-temperature (asymptotic-scaling) regime.
Our novel approach combines a very efficient loop
cluster algorithm -- operating directly in the Euclidean time
continuum -- with finite-size scaling. This enables us to probe
correlation lengths up to $\xi \approx 350,000$ lattice spacings -- more
than three orders of magnitude larger than any previous study.
We resolve a conundrum concerning the applicability of asymptotic-scaling 
formulae to experimentally- and numerically-determined correlation lengths,
and arrive at a very precise determination of the low-energy observables.
Our results have direct implications for the zero-temperature behavior
of spin-1/2 ladders.

\end{abstract}
%

]

\narrowtext

Soon after the discovery of high-temperature superconductivity in doped
lamellar copper oxides it was found that the undoped compounds are
quasi-two-dimensional (2-d) spin $S=1/2$
quantum antiferromagnets. A theoretical model that captures the essential
features of these materials is the nearest-neighbor quantum 
antiferromagnetic Heisenberg model (AFHM) on
a square lattice. Through experimental, numerical, and theoretical efforts much
progress has been made in the understanding of these systems. In particular,
detailed neutron scattering measurements of the spin-spin
correlation length in the magnet $\mbox{Sr}_2\mbox{CuO}_2\mbox{Cl}_2$ were
found to be described quantitatively \cite{Gre94}
by both high-temperature numerical results for the 
AFHM \cite{Mak91} and low-temperature theory for the
renormalized classical regime of the $(2+1)$-d $O(3)$-symmetric non-linear
$\sigma$-model \cite{Cha89,Has91}.

The ground state of the above systems shows long-range antiferromagnetic order,
thus spontaneously breaking the $O(3)$ rotational symmetry to $O(2)$.
The low-energy excitations are two massless bosons called magnons or 
spin-waves. These long-range excitations determine the dynamics at low 
energies. One can use chiral perturbation theory (CPT) to derive universal 
expressions for low-energy observables
in terms of three material-specific parameters: the staggered magnetization
${\cal M}_s$, the spin-wave velocity $c$, and the spin stiffness $\rho_s$
\cite{Leu90}. Both numerical data and predictions of CPT are in
apparent good agreement with experimental results for $S = 1/2$
\cite{Gre94}. However, neutron scattering measurements on 
antiferromagnets with $S > 1/2$ \cite{Gre94,Nak95}
reveal a striking discrepancy with CPT predictions based on 3-loop
asymptotic scaling. It has been suggested that for $S > 1/2$, asymptotic
scaling sets in only at very low temperatures -- that is, for
correlation lengths much larger than those accessed experimentally and
numerically \cite{Els95}.

In this Letter, we investigate the correlation length of the 2-d
nearest-neighbor square-lattice spin-1/2 Heisenberg antiferromagnet at 
unprecedentedly low temperatures. We resolve the puzzle concerning the
applicability of asymptotic scaling to experimentally- and 
numerically-determined correlation lengths \cite{Gre94,Nak95,Els95,Kim97}.
To effect this, we combine an efficient loop cluster algorithm operating 
in the Euclidean time continuum -- hence with zero systematic error --
with a finite-size scaling technique. With this powerful new approach, 
the infinite-volume correlation length $\xi$ is probed up to 
$\approx 350,000$ lattice spacings $a$. This allows us to paint a detailed
picture of the variation of $\xi$ with temperature well into the asymptotic
scaling regime, and to arrive at the presently most precise determination
of the low-energy observables of the spin-1/2 AFHM.
We find that asymptotic scaling of $\xi$ with the 3-loop
$\beta$-function of the 2-d classical $O(3)$ model sets in only at about
$10^5 a$, while a 4-loop fit describes the data already at
$\xi/a \approx 100-200 $, corresponding to the largest correlation lengths 
measured experimentally \cite{Gre94}.
By exchanging a spatial with the Euclidean time
direction, as suggested in Ref.\cite{Cha96}, our results may be 
applied to spin-1/2 ladders at zero temperature.

In CPT, the 2-d quantum spin system is described by a
$(2+1)$-d $O(3)$ symmetric Euclidean field theory. At non-zero temperature $T$
the Euclidean time direction has a finite extent $1/T$. 
For a system of massless particles --
that is, one with an infinite zero-temperature correlation
length -- the non-zero temperature system appears dimensionally reduced to two
dimensions, because $1/T$ is then negligible compared to $\xi$. 
However, the Hohenberg-Mermin-Wagner-Coleman theorem forbids interaction among 
massless Goldstone bosons in two dimensions. Consistent with this, the 2-d 
$O(3)$ model is known to have a non-perturbatively generated mass gap. This 
in turn implies that the spin-waves of a 2-d quantum antiferromagnet at 
non-zero temperature also acquire a mass.
Using perturbative renormalization group arguments, Chakravarty, Halperin, and
Nelson (CHN) \cite{Cha89} derived the expression 
\begin{equation}
\label{CHN}
\xi = 0.31(4) \frac{c}{2 \pi \rho_s} \exp \left( \frac{2 \pi \rho_s}{T} \right)
\left[ 1 + {\cal O} \left( \frac{T}{\rho_s}\right) \right].
\end{equation}
Consistent with dimensional reduction, $\xi$ is exponentially large compared 
to $c/T$. We note that the exponent in Eq.(\ref{CHN})
comes from a 1-loop calculation, while the factor $c/2 \pi \rho_s$ is a
2-loop result.

Hasenfratz and Niedermayer \cite{Has91} averaged the
$(2+1)$-d field over cubic space-time volumes of size $1/T$ in the
Euclidean time direction and $c/T$ in the two spatial directions. 
Since at low temperatures 
$\xi \gg c/T$, the field is essentially constant over these
blocks. The averaged field naturally lives at the block centers, which form a
2-d lattice of spacing $a' = c/T$ (which is different from the
lattice spacing $a$). Hence, the effective action of the
averaged field defines a 2-d lattice $O(3)$ model. Using CPT, as well as the
exact mass-gap and the 3-loop $\beta$-function of the
2-d $O(3)$ model \cite{Has90}, Hasenfratz and Niedermayer extended the
CHN-formula to 
\begin{equation}
\label{CH_2N_2}
\xi = \frac{e}{8} \frac{c}{2 \pi \rho_s}
\exp \left( \frac{2 \pi \rho_s}{T} \right)
\left[ 1 - \frac{T}{4 \pi \rho_s}
+ {\cal O} \left( \frac{T^2}{\rho_s^2} \right) \right],
\end{equation}
which we call the $\mbox{CH}_2\mbox{N}_2$-formula. This equation is valid 
at low temperatures, that is, for large correlation lengths.
When the correlation length is correctly described by Eq.(\ref{CH_2N_2}), it
scales asymptotically with the 3-loop $\beta$-function of the 2-d $O(3)$ model.
The undetermined ${\cal O}(T^2/\rho_s^2)$ term represents a 4-loop effect.

In relativistic quantum field theory the lattice spacing serves as an 
ultraviolet cut-off that is ultimately removed. The question of
asymptotic scaling is thus unphysical because it involves the bare coupling
constant. In the 2-d AFHM, however, the 
lattice spacing of the induced classical model 
is $a' = c/T$, where $T$ is the physical temperature. 
Hence, the question of asymptotic scaling becomes physical. A priori, 
it is unclear for what values of $\xi$ one should expect asymptotic 
scaling for the effective 2-d lattice action.

This raises the important question: for what $\xi$ {\it does} the
$\mbox{CH}_2\mbox{N}_2$-formula work correctly? We address this issue
for the AFHM, defined by the Hamiltonian
\begin{equation}
H = J \sum_{x,\mu} \vec S_x \cdot \vec S_{x+\hat\mu},
\end{equation}
where $J>0$ is the antiferromagnetic coupling, $\vec S_x$ is a quantum 
spin-1/2 operator located at point $x$ of a square lattice with spacing 
$a$, and $\hat\mu$ is the unit vector in the $\mu$-direction.
Previously, comparisons of Monte Carlo calculations
of the staggered and uniform susceptibilities ($\chi_s$ and $\chi_u$) 
with CPT predictions yielded $c = 1.68(1) J a$, $\rho_s = 0.186(4) J$,
and ${\cal M}_s = 0.3083(2) / a^2$
\cite{Wie94,Bea96}. The calculation of Ref.\cite{Wie94} indicates that
CPT results for $\chi_s$ and $\chi_u$ of the same kind as 
Eq.(\ref{CH_2N_2}) -- but with known ${\cal O}(T^2/\rho_s^2)$ terms -- 
work very well for the AFHM for $T \leq 0.2 J$. 
Hence, one expects that asymptotic scaling at the 4-loop level sets 
in at about the same temperature. At $T = 0.2 J$ the 
$\mbox{CH}_2\mbox{N}_2$-formula predicts $\xi/a \approx 150$. 
To avoid finite-size effects in numerical simulations, one must
simulate lattices with a spatial extent at least $L \approx 6 \xi$. Even
with the best algorithms available \cite{Eve93,Wie94,Bea96},
working on lattices starting with $L^2 = 900^2$ is extremely time-consuming.

%

How then can we investigate $\xi$ in the low-temperature
regime? We make use of finite-size effects instead of trying to avoid them. 
Finite-size scaling methods were used by Kim \cite{Kim93} and by 
Caracciolo {\it et al.} \cite{Car95} to test asymptotic scaling of the 
correlation length with the bare coupling constant in the classical 2-d 
lattice $O(3)$ model. To investigate $\xi$ in the AFHM,
we apply the technique of Ref.\cite{Car95}, in which 
renormalization group methods are used to justify the search for a universal 
scaling function 
\begin{equation}
\label{scaling}
\xi(2L) / \xi(L) = F(\xi(L)/L).
\end{equation}
Carracciolo {\it et al.} \cite{Car95} showed that their data indeed collapse 
to a single scaling function $F(\xi(L)/L)$. 
They also showed that iteration of Eq.(\ref{scaling}) yields rapid 
convergence to $\xi$, the correlation length of the infinite system.

%

The universal function $F$ for the classical 2-d $O(3)$ model was 
determined very precisely in Ref.\cite{Car95}. Since at low 
temperatures the quantum model reduces to a classical 2-d lattice $O(3)$ 
model, one can use the same function to deduce 
$\xi$ from $\xi(L)$ data. We verified that indeed the same scaling 
function works for the Heisenberg model with measurements 
of $\xi(L)$ for lattices of successively larger size. This calculation 
yields very good, though not exact, agreement with $F(\xi(L)/L)$. 
We incorporated a tiny correction to the scaling function to account for 
these violations at each stage of the iteration. We found that the
correlation length so determined was insensitive to the 
form of the scaling violation, for several different
fitted functional forms.

%

%

We emphasize that Eq.(\ref{scaling}) assumes universal behavior, that 
is, scaling, but not asymptotic scaling. This is important, because we want
to use Eq.(\ref{scaling}) to compare with the $\mbox{CH}_2\mbox{N}_2$-formula
Eq.(\ref{CH_2N_2}) without bias.

%

The scaling procedure is very sensitive to small changes in the
finite-volume correlation length $\xi(L)$, that is, a small error in 
$\xi(L)$ can lead to large uncertainties in the infinite-volume $\xi$. Hence,
one needs a very accurate numerical method to determine $\xi(L)$. Fortunately, 
for the AFHM there exists a very efficient loop cluster algorithm 
\cite{Eve93,Wie94}, which practically eliminates auto-correlations in 
successive Monte Carlo configurations. The cluster algorithm also enables
improved estimators which drastically reduce statistical errors. Finally, 
implementing the cluster algorithm in continuous Euclidean time completely 
eliminates the systematic error due to the Trotter-Suzuki discretization of 
time \cite{Bea96}. This continuous-time cluster algorithm (CTCA) has greatly 
reduced storage and computer time requirements, enabling simulation at very 
low temperatures.

\begin{figure}
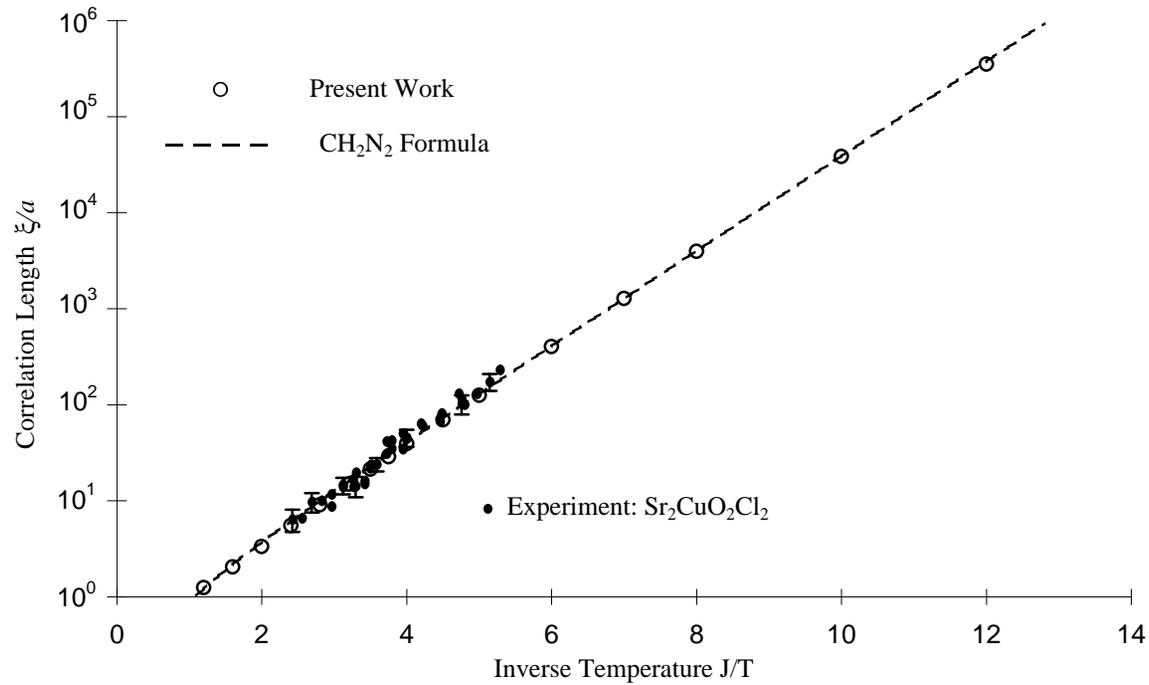

\caption{Experimental \protect\cite{Gre94}
(filled circles) and computed values (open circles;
errors are much smaller than symbol size)
of the staggered-spin correlation length.
The solid line is the 3-loop $\mbox{CH}_2\mbox{N}_2$-formula, 
Eq.(\protect\ref{CH_2N_2}), 
with $\rho_s=0.1800$.}
\label{logchart}
\end{figure}

For $J/T = 0.5,1.0,...,3.5$, we carry out simulations that
measure the correlation length directly in a large volume $L \approx 6 \xi$,
while for $J/T = 4.0,4.5,...,12.0$, we use the finite-size scaling technique.
A single-cluster version of the CTCA is used with an improved estimator 
for the staggered correlation function, then $\xi(L)$ is extracted from the 
correlation function using the second-moment method \cite{Kim93,Car95}. 
For the direct-measurement cases, $10^5$ Monte Carlo configurations are 
used; for the scaled-measurement cases, $4 \times 10^5$ are used. The 
numerical data for $\xi(L)$ together with the inferred infinite-volume 
$\xi$ are given in Table \ref{xi_vs_beta}, and are compared with experimental
data and the $\mbox{CH}_2\mbox{N}_2$-formula in Fig.\ref{logchart}. The 
largest accessible $\xi$ is more than three orders of magnitude larger than 
those of any previous study \cite{Gre94,Mak91,Gre96,Kim97}.

\begin{figure}
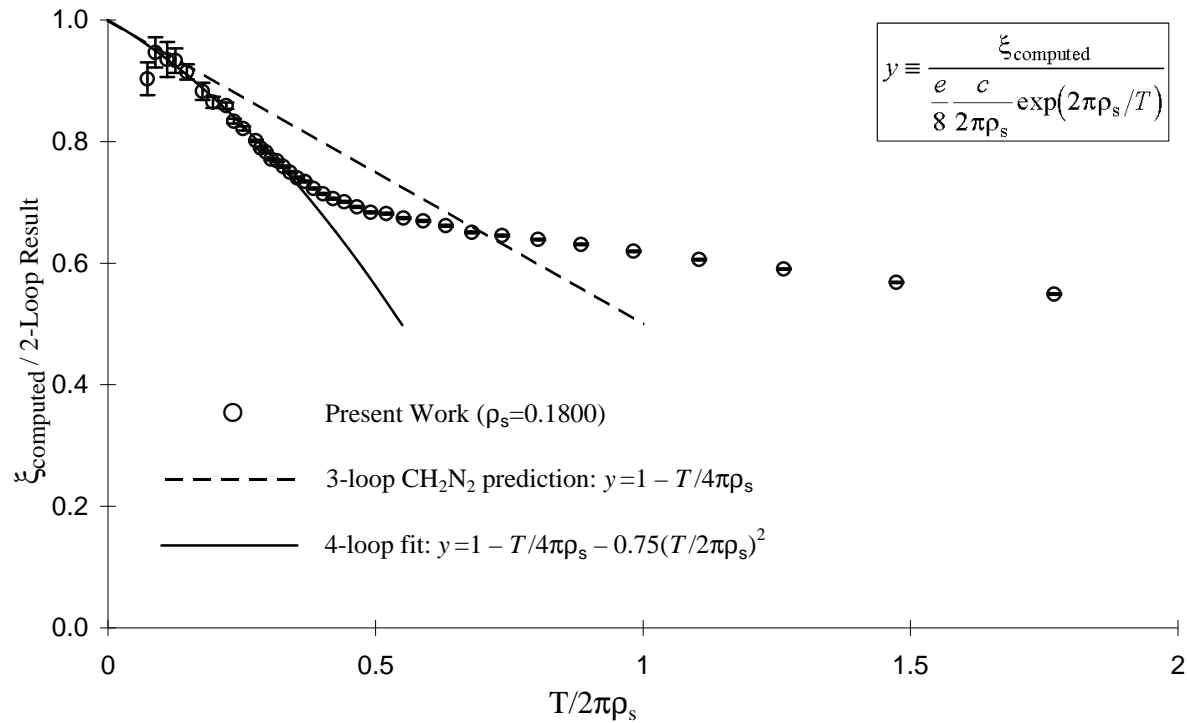

\caption{Analysis of the correlation length as the asymptotic scaling regime
is approached.
To remove the exponential inverse-temperature dependence,
$\xi$ is divided by the 2-loop formula,
using the best-fit value $\rho_s=0.1800(5)$.
The 3-loop result (straight line) is accurate only for $\xi/a \approx 10^5$,
while the 4-loop regime (parabola) begins at roughly $\xi/a \approx 10^2$.
}
\label{crisischart}
\end{figure}

Deviations from asymptotic scaling are invisible on the semi-log scale of 
Fig.\ref{logchart}. Figure \ref{crisischart} shows the deviation from
2-loop asymptotic scaling as a function of $T/2 \pi \rho_s$.
Here, the 3-loop result of Eq.(\ref{CH_2N_2}) is a line of slope $-1/2$.
The quantitative agreement shown in Fig.\ref{logchart} between experiment 
and theory in the regime $2.4 < J/T < 5.3$ is seen to be 
coincidental. We find that with decreasing temperature the correlation 
length crosses and undershoots the 3-loop result before approaching it from 
below. Consequently, the deviations at intermediate and high temperatures 
from the $\mbox{CH}_2\mbox{N}_2$-formula turn out to be relatively small.
Although neutron scattering results for $\rm Sr_2 Cu O_2 Cl_2$
cover a wide range of $\xi$ with relatively small errors 
\cite{Gre94}, the deviations from 3-loop asymptotic scaling turn out to be
too subtle to be discernable experimentally. This explains why the 
low-temperature result of Eq.(\ref{CH_2N_2}) appears to
describe the experimental data at such high temperatures.

Owing to the exponential dependence on $\rho_s$, the placement of the 
numerical data on the graph of Fig.\ref{crisischart} is extremely 
sensitive to $\rho_s$ in the low-$T$ regime. This enables a very 
precise estimate of $\rho_s$. The correlation length data
are fitted simultaneously with previously obtained results for 
$\chi_s$ and $\chi_u$ in cubical \cite{Wie94} and cylindrical
\cite{Bea96} space-time geometries,
resulting in
\begin{eqnarray*}
c = 1.657(2) J a, \ \rho_s = 0.1800(5) J, \ {\cal M}_s = 0.30797(3) / a^2 .
\end{eqnarray*}
This is in very good agreement with results of an expansion around the
Ising limit, which gives 
$c = 1.654(11) J a$,
$\rho_s = 0.182(5) J$, and 
${\cal M}_s = 0.307(1) / a^2$ \cite{Wei91}.

In order to fit the correlation length data down to
$\xi/a \approx 100$, it was necessary to include the ${\cal O}(T^2/\rho_s^2)$
4-loop term -- not determined in Ref.\cite{Has90} --
with a fitted coefficient $C_2$.
We find $C_2 = -0.75(5)$ if the fit is restricted to
quadratic order and to $J/T > 2.8$.
Higher-order fits (including $C_3$, $C_4$, ...)
shift the value determined for $C_2$, although it remains 
${\cal O}(1)$ and stable.
A previous study also found that quadratic-order terms are needed to
match the precision of the $\chi_s$ and $\chi_u$ measured
with the CTCA \cite{Bea96}.
Thus, it should come as no surprise that asymptotic scaling at the 3-loop
level of the $\mbox{CH}_2\mbox{N}_2$-formula sets in at rather large
correlation lengths of $\xi \approx 10^5 a$. The same is true for the 
classical 2-d $O(3)$ model with the standard lattice action \cite{Car95}.

%

We also carried out an additional independent fit for the prefactor 
in the $\mbox{CH}_2\mbox{N}_2$-formula. That is, 
we multiplied Eq.(\ref{CH_2N_2}) by an extra fitted parameter $A$,
and found $A=1.02(2)$ (using the 4-loop form for the $\beta$-function),
in excellent agreement with Eq.(\ref{CH_2N_2}). This gives added 
confidence both in our scaling procedure and in the correctness
of the exact massgap. 

%

A large discrepancy between the $\mbox{CH}_2\mbox{N}_2$-formula and
experimental data has been discovered for systems with $S > 1/2$
\cite{Gre94,Nak95}. A likely explanation of this discrepancy is that
3-loop asymptotic scaling again sets in only at very small temperatures. In
fact, the existence of a crossover from high- to low-temperature behavior was
suggested in Ref.\cite{Els95}. Although the asymptotic region is inaccessible
to experiments, we have demonstrated that it is possible to investigate it
using the cluster algorithm in combination with finite-size scaling.
A similar study for $S > 1/2$ is presently in progress.

Recently, it was pointed out that the $\mbox{CH}_2\mbox{N}_2$-formula can
also be applied to the correlation length at $T=0$ of even-width Heisenberg
spin-ladders \cite{Cha96}. These ladders comprise an even number $n$
of coupled $S=1/2$ chains with periodic boundary conditions. This mapping 
exchanges one spatial direction with the Euclidean time direction, so the 
time extent plays the role of $n$. The 3-loop $\beta$-function is the 
same for the square-lattice and the ladder. In contrast, the undetermined 
higher-order terms are influenced by the distinction that space is a 
lattice, whereas Euclidean time is continuous. Hence, 
the ${\cal O}(T^2/\rho_s^2)$ terms are expected to be different for the 
square-lattice and the spin-ladders.
%
Early density-matrix renormalization group calculations of the spin gap of 
even-width ladders were reported in Ref.\cite{Whi94}.
%
The correlation length of spin-1/2 ladders was studied numerically 
for $n \leq 6$ \cite{Gre96,Syl97}. We see in Fig.\ref{crisischart} 
that 3-loop asymptotic scaling sets in only at $T \approx 0.13 J$. 
We thus expect the 3-loop $\mbox{CH}_2\mbox{N}_2$-formula 
to work quantitatively only for ladders with widths greater than 
$n = c/T \approx 12$. It was already realized in Ref.\cite{Syl97} that 
the 3-loop result is insufficient to describe the numerical data 
for $n \leq 6$. A 4-loop form should work for $n \geq 6$.

In conclusion, we have combined a powerful and
accurate quantum Monte Carlo technique with finite-size scaling.
This has made possible the determination of the
correlation length of the $S = 1/2$ nearest-neighbor
square-lattice antiferromagnetic
Heisenberg model at unprecedentedly low temperatures.
Our study resolves the conundrum concerning the applicability
of the 3-loop asymptotic scaling description for $S = 1/2$,
and it points the way toward a quantitative resolution of this issue
for $S > 1/2$.
Our results also have direct implications for the
low-temperature properties
of $S = 1/2$ ladders.

We are indebted to A. Ferrando and F. Niedermayer for very interesting 
discussions. We also thank A. Pelissetto for providing us with the 
universal finite-size-scaling function $F$. This work was supported 
by the NSF under Grant No. DMR 97-04532, the International Joint 
Research Program of NEDO (New Energy Development Organization)
International Joint Research Grant, and the DOE under cooperative 
research agreement number DE-FC02-94ER40818. U.-J.W. is
also supported by an A.P. Sloan fellowship.

%
\begin{table}
\caption{Finite- and infinite-volume correlation length. 
Infinite-volume $\xi/a$ are deduced using finite-size scaling.}
\begin{tabular}{r@{}c@{}l c r@{}c@{}l r@{}c@{}l@{}}
  \multicolumn{3}{c}{$J/T$} & $L/a$ & \multicolumn{3}{c}{$\xi(L)/a$}
  & \multicolumn{3}{c}{$\xi/a \equiv \xi(\infty)/a$} \\
\tableline
 0&.&5  &       20  &        0&.&481(1) &        0&.&481(1) \\
 1&.&0  &       80  &        0&.&973(2) &        0&.&973(2) \\
 1&.&5  &       80  &        1&.&818(4) &        1&.&818(4) \\
 2&.&0  &       80  &        3&.&351(8) &        3&.&351(8) \\
 2&.&5  &       80  &        6&.&23(1)  &        6&.&23(1)  \\
 3&.&0  &      160  &       11&.&60(3)  &       11&.&60(3)  \\
 3&.&5  &      160  &       21&.&23(2)  &       21&.&25(2)  \\
 4&.&0  &       80  &       31&.&3(1)   &       39&.&4(2)   \\
 4&.&5  &       80  &       39&.&0(1)   &       69&.&9(7)   \\
 5&.&0  &       80  &       44&.&6(1)   &      126&&(2)     \\
 6&.&0  &       80  &       52&.&5(1)   &      403&&(6)     \\
 7&.&0  &       80  &       58&.&8(1)   &     1275&&(27)    \\
 8&.&0  &       80  &       64&.&2(1)   &     3960&&(120)   \\
10&.&0  &       80  &       73&.&5(1)   &    38460&&(1010)  \\
12&.&0  &      160  &      159&.&0(2)   &   352000&&(10500) \\
\end{tabular}
\label{xi_vs_beta}
\end{table}


\newpage

\begin{references}

\bibitem{Gre94}
M. Greven {\it et al.}, Phys. Rev. Lett. {\bf 72}, 1096 (1994);
Z. Phys. B {\bf 96}, 465 (1995).

\bibitem{Mak91}
M.S. Makivi\'c and H.-Q. Ding, Phys. Rev. B {\bf 43}, 3562 (1991).

\bibitem{Cha89}
S. Chakravarty, B.I. Halperin, and D.R. Nelson, Phys. Rev. B {\bf 39}, 2344
(1989).

\bibitem{Has91}
P. Hasenfratz and F. Niedermayer, Phys. Lett. B {\bf 268}, 231 (1991).

\bibitem{Leu90}
P. Hasenfratz and H. Leutwyler, Nucl. Phys. B {\bf 343}, 241 (1990);
P. Hasenfratz and F. Niedermayer, Z. Phys. B {\bf 92}, 91 (1993).

\bibitem{Nak95}
K. Nakajima {\it et al.}, Z. Phys. B {\bf 96}, 479 (1995);
Y.S. Lee {\it et al.}, to be published.

\bibitem{Els95}
N. Elstner {\it et al.},
Phys. Rev. Lett. {\bf 75}, 938 (1995).

\bibitem{Kim97}
J.-K. Kim, D.P. Landau, and M. Troyer, Phys. Rev. Lett. {\bf 79}, 1583 (1997).

\bibitem{Cha96}
S. Chakravarty, Phys. Rev. Lett. {\bf 77}, 4446 (1996).

\bibitem{Has90}
P. Hasenfratz, M. Maggiore, and F. Niedermayer, Phys. Lett. B {\bf 245}, 522
(1990); P. Hasenfratz and F. Niedermayer, Phys. Lett. B {\bf 245}, 529 (1990).

\bibitem{Wie94}
U.-J. Wiese and H.-P. Ying, Z. Phys. B {\bf 93}, 147 (1994).

\bibitem{Bea96}
B.B. Beard and U.-J. Wiese, Phys. Rev. Lett. {\bf 77}, 5130 (1996).

\bibitem{Eve93}
H.G. Evertz, G. Lana, and M. Marcu, Phys. Rev. Lett. {\bf 70}, 875 (1993).

\bibitem{Kim93}
J. K. Kim, Phys. Rev. Lett. {\bf 70}, 1735 (1993); Phys. Rev. D {\bf 50}, 4663
(1994).

\bibitem{Car95}
S. Caracciolo {\it et al.}, Phys. Rev. Lett. {\bf 75}, 1891 (1995).

\bibitem{Gre96}
M. Greven, U.-J. Wiese, and R.J. Birgeneau,
Phys. Rev. Lett. {\bf 77}, 1865 (1996).

\bibitem{Wei91}
Z. Weihong, J. Oitmaa, and C.J. Hamer, Phys. Rev. B {\bf 43}, 8321 (1991);
C. Hamer, Z. Weihong, and J. Oitmaa, Phys. Rev. B {\bf 50}, 6877 (1994).

%
\bibitem{Whi94}
S.R. White, R.M. Noack, and D.J. Scalapino, 
Phys. Rev. Lett. {\bf 73}, 886 (1994). 
%

\bibitem{Syl97}
O.F. Sylju{\aa}sen, S. Chakravarty, and M. Greven, 
Phys. Rev. Lett. {\bf 78}, 4115 (1997).

\end{references}
\end{document}